\documentclass[aps,pre,twocolumn,superscriptaddress]{revtex4-1}
\usepackage{feynmf}
\usepackage{graphicx}
\usepackage{bm}
\usepackage{url}
\usepackage{physics}
\usepackage{color}
\usepackage{amsmath}
\usepackage{enumerate}
\usepackage{amsfonts}
\usepackage{epsfig}
\usepackage{cancel}
\usepackage{bbm}
\usepackage{amsfonts}
\usepackage{threeparttable}
\usepackage{float}
\usepackage[caption=false]{subfig}
\usepackage{hyperref}

\DeclareMathOperator{\sign}{sgn}

\newcommand{\subfigimg}[3][,]{%
  \setbox1=\hbox{\includegraphics[#1]{#3}}
  \leavevmode\rlap{\usebox1}
  \rlap{\raisebox{\dimexpr\ht1-1\baselineskip}{\footnotesize(#2)}}
  \phantom{\usebox1}
}

\newcommand{\angles}[1]{%
\left \langle #1 \right \rangle%
}

\begin{document}


\title{Dynamics around the Site Percolation Threshold on High-Dimensional Hypercubic Lattices}



\author{Giulio Biroli}
\affiliation{Laboratoire de Physique Statistique, Ecole Normale Sup\'erieure,
PSL Research University, 24 rue Lhomond, 75005 Paris, France}
\author{Patrick Charbonneau}
\affiliation{Department of Chemistry, Duke University, Durham, NC 27708, USA}
\affiliation{Department of Physics, Duke University, Durham, NC 27708, USA}
\author{Yi Hu}
\affiliation{Department of Chemistry, Duke University, Durham, NC 27708, USA}

\date{\today}

\begin{abstract}
Recent advances on the glass problem motivate reexamining classical models of percolation. Here, we consider the displacement of an ant in a labyrinth near the percolation threshold on cubic lattices both below and above the upper critical dimension of simple percolation, $d_\mathrm{u}=6$. Using theory and simulations, we consider the scaling regime part, and obtain that both caging and subdiffusion scale logarithmically for $d \ge d_\mathrm{u}$. 
The theoretical derivation considers Bethe lattices with generalized connectivity and a random graph model, and employs a scaling analysis to confirm that logarithmic scalings should persist in the limit $d\rightarrow\infty$. The computational validation employs accelerated random walk simulations with a transfer-matrix description of diffusion to evaluate directly the dynamical critical exponents below $d_\mathrm{u}$ as well as their logarithmic scaling above $d_\mathrm{u}$. Our numerical results improve various earlier estimates and are fully consistent with our theoretical predictions.
\end{abstract}


\maketitle

\section{Introduction}
Transport in disordered media is anomalous compared to its counterpart in homogeneous space~\cite{kirkpatrick1973percolation,bouchaud1990anomalous,ben2000diffusion}. Diffusion in systems as diverse as porous rocks, aerogels and biological cells is indeed much more complex than Einstein's description of Brownian motion~\cite{ben2000diffusion,hofling2013anomalous}. The paradigmatic minimal model for such transport is de Gennes' ant in a labyrinth~\cite{gennes1976percolation}, which consists of randomly displacing a tracer on covered lattice sites, around the percolation threshold. While, far above that threshold, transport is unremarkable---other than being more sluggish than an unimpeded random walk---near the threshold a clear subdiffusive regime emerges, and below the threshold transport stops altogether.

Most features of this model have by now been extensively studied by theory and simulations. Scaling relations, series expansions, and renormalization group treatments are very well developed, and most critical exponents are known with high precision~\cite{stauffer1994percolation}. Many aspects of the process have even been fully mathematically formalized~\cite{heydenreich2017progress,arous2016scaling,arous2016scalingb,arous2017scaling}. Yet some of its features remain actively pursued, thus reflecting the continued importance and elegance of the underlying physical model. Improved numerical estimates of the thresholds, for instance, keep appearing~\cite{xu2014simultaneous,mertens2018percolation}. Careful studies of the localized regime, which has traditionally been less studied than the conduction side, are also now emerging, especially on higher-dimensional lattices. Mertens and Moore, for instance, recently improved the high-dimensional series expansion for the threshold from a finite-cluster expansion~\cite{mertens2018series}, and employed specialized simulation techniques to compute more precise Fisher exponents, which describe the large cluster size distribution~\cite{mertens2018percolation}. 

Recent advances in the field of glasses also motivate reexamining the subdiffusive and localized regime of the ant in a labyrinth. The study of the Mari-Kurchan model, which offers a finite-dimensional mean-field description of glasses in continuous space~\cite{mari2009jamming,mari2011dynamical,mezard2011solution}, suggests that a percolation-like process might play a role in the dynamical slowing down in finite $d$~\cite{charbonneau2014hopping,jin2015dimensional}. An interesting common feature of Mari-Kurchan glass formers and ants in labyrinths is that the power-law scaling of subdiffusion could vanish in high spatial dimension, $d$, for both. In the former, true caging actually is expected at the dynamical transition in the limit $d\rightarrow\infty$. In the latter, however, the situation is more ambiguous. While the power-law exponent of subdiffusion is expected to vanish approaching the upper critical dimension for simple percolation, $d_\mathrm{u}=6$, what then follows is unclear. In particular, is caging taking place? Or, does subdiffusion rather become slower than power law?

In this article, we provide an answer to these questions and present a direct evaluation of the critical exponents associated with caging and with subdiffusion from $d=2$ to $d=8$. Although caging exponents had not been previously evaluated directly, our results are fully consistent with the relevant scaling relations. The subdiffusion exponents we obtain are also consistent with scaling theory predictions, but are one order of magnitude more accurate than previous numerical estimates in $d=3$ and $4$. We also derive the logarithmic scaling behavior for Bethe lattices of arbitrary connectivity and for random graphs, and observe the corresponding scaling form on hypercubic lattices with $d \ge d_\mathrm{u}$. The rest of the paper is organized as follows. In Section \ref{sec:theory}, we generalize both the scaling analysis that relate caging and dynamical exponents and the description of caging on Bethe lattices and on random graphs. In Section~\ref{sec:simulations}, we describe the computational scheme used to evaluate these predictions. Section \ref{sec:results} presents and discusses the numerical results for $d=2$ to 8, and we briefly conclude in Section~\ref{sec:conclusion}.

\section{Theoretical Framework}
\label{sec:theory}
In this section, we obtain relations between three dynamical exponents by scaling analysis below the upper critical dimension, $d_\mathrm{u}$, and  discuss the case $d \ge d_\mathrm{u}$. We also compute the critical behavior for caging on a Bethe lattice of arbitrary connectivity and separately consider the same problem on a random graph, which recovers the fully-connected limit of the Bethe lattice.

\subsection{Scaling Analysis}
In order to investigate the dynamical exponents around the site percolation threshold, $p_\mathrm{c}$, we consider the mean-square displacement of the tracer, $D^2(t,p)=\langle r^2(t)\rangle$, at covering fraction $p$ and time $t$. On general grounds~\cite{ben2000diffusion}, we expect the following scaling forms to be obeyed
\begin{equation} \label{eq:sclform}
D^2(t\rightarrow\infty,p) \sim  
\begin{cases}
(p_\mathrm{c}-p)^{-\mu_-}, & p < p_\mathrm{c} \\
t (p-p_\mathrm{c})^{\mu}, & p > p_\mathrm{c} \\
t^{2/d_\mathrm{w}'}, & p = p_\mathrm{c}
\end{cases},
\end{equation}
where the exponent $\mu$ characterizes the decay of the diffusivity upon approaching $p_\mathrm{c}$ from above, $\mu_-$ is the caging exponent that characterizes the growth of the infinite-time limit of the mean-square displacement, $\lim_{t\rightarrow\infty}D^2(p<p_\mathrm{c},t)\equiv\Delta^2(p)$, upon approaching $p_\mathrm{c}$ from below, and $d'_\mathrm{w}$ describes the subdiffusive scaling.
This set of relationships gives rise to the following (non-unique) scaling collapse for $\epsilon\equiv (p-p_\mathrm{c})/p_\mathrm{c}$
\begin{align}
D^2(t,p)&=|\epsilon|^{-\mu_-} f(\sign(\epsilon)|\epsilon|^{\mu_\mathrm{s}}t),
\end{align}
along with three relations. First, caging for $p<p_\mathrm{c}$ corresponds to
\begin{equation*}
\lim_{x\rightarrow-\infty} f(x)=\mathrm{cnst}.
\end{equation*}
with a negative argument to $f(x)$. Second, diffusion for $p>p_\mathrm{c}$ is recovered if 
\begin{equation*}
\lim_{x\rightarrow\infty} f(x)=x,
\end{equation*}
and thus $t\epsilon^\mu=t\epsilon^{\mu_\mathrm{s}-\mu_-}$.
Third, at (reasonably) short times subdiffusion is recovered if
\begin{equation*}
f(x\rightarrow0)=x^{2/{d_\mathrm{w}'}},
\end{equation*}
or, equivalently,
\begin{equation}
D^2(t,p)\sim|\epsilon|^{-\mu_-}(|\epsilon|^{\mu_\mathrm{s}}t)^{2/{d_\mathrm{w}'}}.
\end{equation}
Because the subdiffusive scaling should be independent of $\epsilon$, we must have $\mu_-=\frac{2}{d_\mathrm{w}'}\mu_\mathrm{s}$. Altogether, we therefore obtain
\begin{align}
\mu_\mathrm{s}&=\mu+\mu_-\\
d_\mathrm{w}'&=2 \left(\frac{\mu}{\mu_-}+1\right).\label{eq:dwp}
\end{align}

Based on the known critical scaling relations~\cite{stauffer1994percolation,ben2000diffusion}, one can also compute the caging exponent $\mu_-$ as~\cite{stauffer1994percolation}
\begin{align} \label{eq:muminus}
  \mu_- &= 2\nu - \beta \\ \label{eq:muminus_2}
    &= 2\frac{\tau-1}{\sigma d} - \frac{\tau -2}{\sigma},
\end{align}       
and the subdiffusion exponent as
\begin{equation} \label{eq:dwprime}
d'_\mathrm{w} = 2\left( \frac{\mu}{2\nu - \beta} + 1\right)
\end{equation}
where $\nu$ is the correlation length exponent, $\beta$ is the exponent that characterizes the fraction of infinite network sites, $\tau$ is the Fischer exponent for the cluster size distribution at $p_\mathrm{c}$, and $\sigma$ is the exponent that characterizes the scaling of the large cluster cutoff for $p<p_\mathrm{c}$.

Interestingly, for $d \ge d_\mathrm{u}=6$, which is the upper critical dimension for simple percolation, these scaling relations suggest that $\mu_-=0$  and $d'_\mathrm{w} \rightarrow \infty$. The subdiffusive regime is then either
slower than any power law or fully arrested, as would be a glass former beyond the dynamical transition~\cite{charbonneau2017glass}.  In order to settle the issue and, especially, working out the percolation behavior in the limit $d\rightarrow\infty$, a more detailed treatment is needed. This is the focus of the next subsections.

\begin{table*}
\centering
\begin{threeparttable}
\caption{Reference percolation thresholds and critical exponents used in this work}
\begin{tabular}{ccccccc}
\hline \hline
Dimension & $p_\mathrm{c}$~\cite{mertens2018percolation} & $\nu$~\cite{koza2016discrete} & $\beta$~\cite{adler1990low} & $\mu^\dagger$ & $\tau$~\cite{mertens2018percolation} & $\sigma^{\ddagger}$ \\
\hline
3 & 0.3116077(4)~\cite{deng2005monte} & 0.8774(13) & 0.405(25) & 2.00(2) & 2.1892(1) & 0.4522(8)~\cite{ballesteros1999scaling} \\
4 & 0.19688561(3) & 0.6852(28) & 0.639(20) & 2.42(2) & 2.3142(5) & 0.4742~\cite{gracey2015four} \\
5 & 0.14079633(4) & 0.5723(18) & 0.835(5) & 2.74(2) & 2.419(1) & 0.4933~\cite{gracey2015four} \\
6 & 0.109016661(8) & 1/2 & 1 & 3 & 5/2 & 1/2 \\
7 & 0.088951121(1) & 1/2 & 1 & 3 & 5/2 & 1/2 \\
8 & 0.075210128(1) & 1/2 & 1 & 3 & 5/2 & 1/2 \\
\hline
\end{tabular}
\begin{tablenotes}
      \small
      \item[$\dagger$] Computed from the scaling relation, $\mu= \tilde{\zeta}_R+\nu(d-2)$, with $\tilde{\zeta}_R$ from Ref.~\cite{fisch1978critical} and $\nu$ from this table.
      \item[$\ddagger$] Numerical result for $d=3$ and renormalization group (RG) prediction for $d=4$ and 5. Existing RG predictions for $d=3$ lie outside of the numerical error bars.
\end{tablenotes} \label{tab:refdata}
\end{threeparttable}
\end{table*}

\subsection{Upper critical dimension and above}
Based on the standard theory of critical phenomena, the critical behavior in $d \ge d_\mathrm{u}=6$ is expected to be mean-field like and independent of $d$. As discussed above, because $2/d'_\mathrm{w}$ vanishes for $d\rightarrow d_\mathrm{u}^-$, however, it is not clear whether the subdiffusive regime is then slower than any power law or fully arrested. In order to clarify this point, we briefly recall the physics behind the above scaling relations. 

Consider the scaling behavior at $p=p_\mathrm{c}$. Percolation theory~\cite{stauffer1994percolation} indicates that a tracer belongs to a cluster of size $s$ with probability $P(s)\sim s^{1-\tau}$. On the infinite cluster---as well as on a large cluster---a tracer subdiffuses with a scaling law $r(t)\sim t^{1/d_\mathrm{w}}$, which is distinct from $d'_\mathrm{w}$. Given that  
a cluster of linear size $\ell$ and fractal dimension $d_\mathrm{f}$ contains $s\sim\ell^{d_\mathrm{f}}$ sites, after a time $t$
a tracer can fully explore clusters of size $s<s^*\sim t^{d_\mathrm{f}/d_\mathrm{w}}$. 
For tracers on larger clusters, only part of the available sites can be explored. To compute the mean-square displacement one should thus consider two contributions: one for a tracer on a cluster with $s<s^*$, and the other for clusters with $s \ge s^*$. The mean-square displacement then reads
\begin{equation}\label{eq:pl}
\begin{aligned}
D^2(t\rightarrow \infty,p_\mathrm{c}) &\sim \int_1^{s^*} \dd s P(s) s^{2/d_\mathrm{f}} + \int_{s^*}^\infty \dd s P(s) t^{2/d_\mathrm{w}} \\
&\sim (s^*)^{2-\tau+2/d_\mathrm{f}} + (s^*)^{2-\tau} t^{2/d_\mathrm{w}} \\
&\sim t^{\frac{d_\mathrm{f}}{d_\mathrm{w}}(2-\tau + \frac{2}{d_\mathrm{f}})},
\end{aligned}
\end{equation}
where in the last step we replaced $s^*$ with $t^{d_\mathrm{f}/d_\mathrm{w}}$. Note that because the two contributions grow similarly, we can consider the scaling of either for our analysis.

Using scaling relations, one also gets
\begin{equation} 
\frac{2}{d'_\mathrm{w}} = \frac{d_\mathrm{f}}{d_\mathrm{w}} (2-\tau+2/d_\mathrm{f})
 = \frac{2-\beta/\nu}{d_\mathrm{w}}.
\end{equation}
Given that $d_\mathrm{f}\rightarrow 4$, $d_\mathrm{w}\rightarrow 6$ and $(2-\beta/\nu)\rightarrow 0$ upon approaching $d_\mathrm{u}$ from below~\cite{ben2000diffusion}, one finds that 
the subdiffusion exponent vanishes, which is the result quoted above. In this case, one should compute the integral in Eq.~\eqref{eq:pl} as
\begin{equation} \label{eq:plln}
\begin{aligned}
D^2(t\rightarrow \infty,p_\mathrm{c}) &\sim \int_1^{s^*} \dd s P(s) s^{2/d_\mathrm{f}} + \int_{s^*}^\infty \dd s P(s) t^{2/d_\mathrm{w}} \\
&\sim \int_1^{s^*} s^{-1} \dd s + (s^*)^{2-\tau} t^{2/d_\mathrm{w}} \\
&\sim \ln s^* + \mathrm{cnst} \\
&\approx A_D \ln t,
\end{aligned} 
\end{equation}
where we explicitly denote the prefactor as a dimensionally-dependent constant, $A_D(d)$. The scaling analysis thus predicts for $d\geq d_\mathrm{u}$ that $D^2(t, p_c)$ grows logarithmically with time.
One can similarly obtain the scaling behavior of caging
for $p<p_\mathrm{c}$. Because the power-law scaling of $P(s)$ is cut off at $s^*\sim \epsilon ^{-1/\sigma}$, 
the singular contribution to the long-time mean-square displacement can again be obtained from Eq.~\eqref{eq:pl}, after replacing 
$s^*$ with $\epsilon ^{-1/\sigma}$. This leads to $\mu_-=2\nu-\beta$, which vanishes for $d\rightarrow d_\mathrm{u}^-$. Integrating similarly gives $\Delta^2 = - A_\Delta \ln |\epsilon|$, with the prefactors $A_\Delta$ and $A_D$ satisfying the relation
\begin{equation*} 
\begin{aligned}
&D^2(t\rightarrow \infty,p_\mathrm{c}) = -A_\Delta \ln |\epsilon| = A_D \ln t \\
\Rightarrow &-A_\Delta \ln \left[ (s^*)^{-\sigma}\right] = A_D \ln \left[ (s^*)^{d_\mathrm{w}/{d_\mathrm{f}}} \right], \\
\end{aligned} 
\end{equation*}
and thus
\begin{equation} \label{eq:prefrelation}
A_\Delta = \frac{d_\mathrm{w}}{\sigma d_\mathrm{f}} A_D.
\end{equation}
Inserting the critical exponents for $d\geq d_\mathrm{u}$ immediately gives $A_\Delta = 3 A_D$.

This derivation pinpoints the origin of the phenomenon we wish to understand. That is, subdiffusion disappears because the exponent relating the mean-square displacement to $s^*(t)$ vanishes, whereas 
the time-dependence of $s^*(t)$ remains a power law. One indeed expects that above 
the upper critical dimension $s^*(t)$ scales as $t^{d_\mathrm{f}/d_\mathrm{w}}=t^{2/3}$. Analogously, the dependence of $s^*$ on $\epsilon$ is $\epsilon^{-1/\sigma}$ with 
$\sigma=1/2$ for all $d>d_\mathrm{u}$.

Mertens and Moore further suggest the Fischer exponent in $d_\mathrm{u} = 6$ combines with a logarithmic correction~\cite{mertens2018percolation}, as suggested by early renormalization group treatments~\cite{essam1978percolation,nakanishi1980scaling}. 
Under this correction, we would have
\begin{equation}
P(s) \sim s^{1-\tau} (\ln s)^\theta,
\end{equation}
and Eq.~\eqref{eq:plln} then becomes
\begin{equation} \label{eq:6dDscaling}
\begin{aligned}
D^2(t\rightarrow \infty,p_\mathrm{c}) &\sim \int_1^{s^*} s^{-1} (\ln s)^\theta \dd s \\
&\approx A_D (\ln t)^{1+\theta},
\end{aligned} 
\end{equation}
and similarly
\begin{equation} \label{eq:6dscaling}
\Delta^2 \approx A_\Delta (-\ln |\epsilon|)^{1+\theta},
\end{equation}
with $A_\Delta = 3^{1+\theta} A_D$. The logarithmic correction should then be observed in both the scaling of the caging and of the dynamics at $p_\mathrm{c}$.

The above analyses, however, only predict a logarithmic scaling in $d \ge d_\mathrm{u}$, not how its prefactor evolves with $d$, and especially whether this prefactor remains finite in the limit $d\rightarrow\infty$. This question can only be resolved by computing explicitly the full dependence on $\epsilon$, and hence on $s^*$, for $p<p_\mathrm{c}$, which we do for mean-field lattices in the following two subsections.

\subsection{Tracer on a Bethe lattice} \label{sec:bethe}
\begin{figure}[!htbp]
  \centering
  \subfloat{\subfigimg[width=0.45\columnwidth]{a}{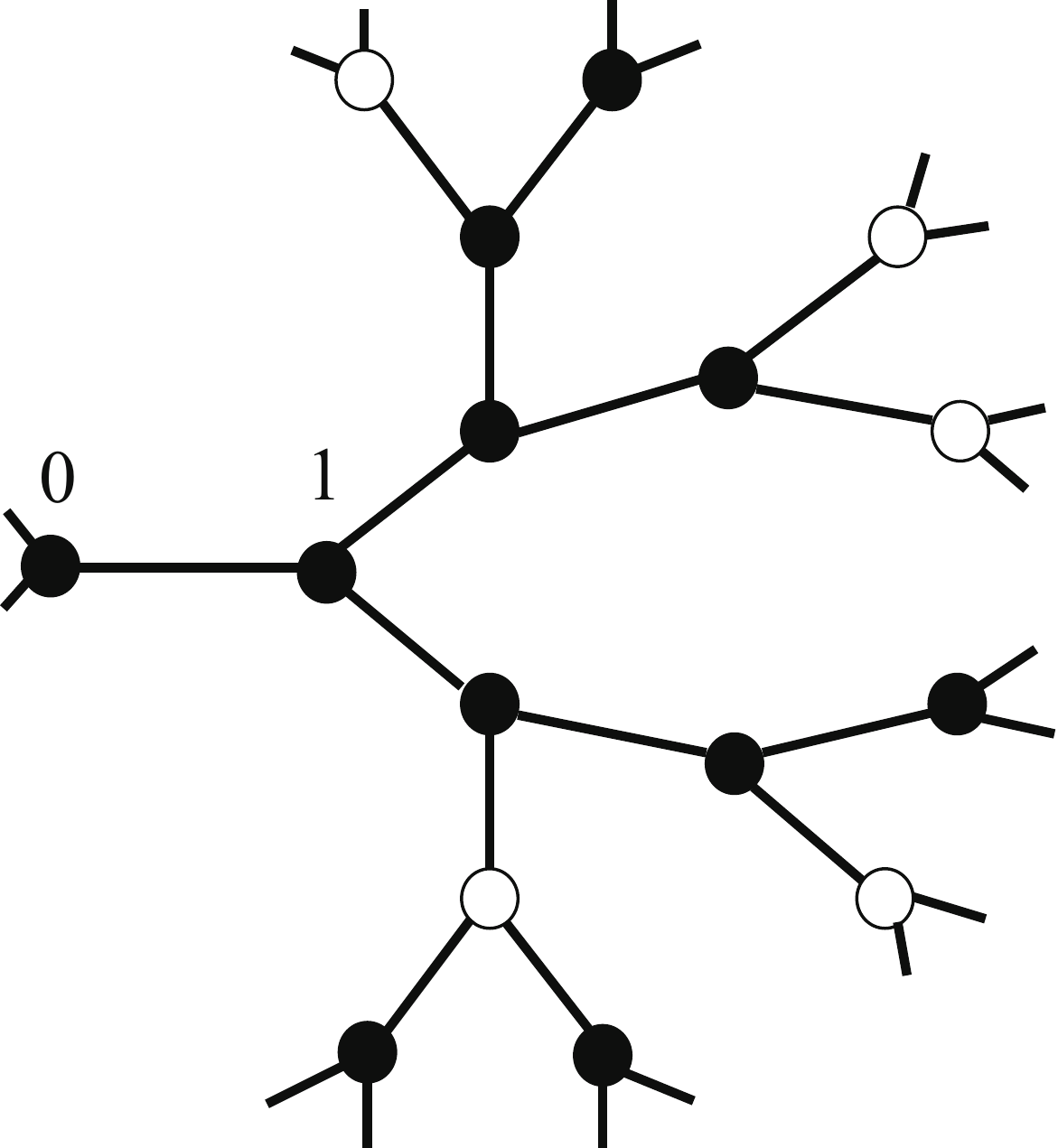}} 
  \subfloat{\subfigimg[trim={-20 -100 0 0},clip,width=0.45\columnwidth]{b}{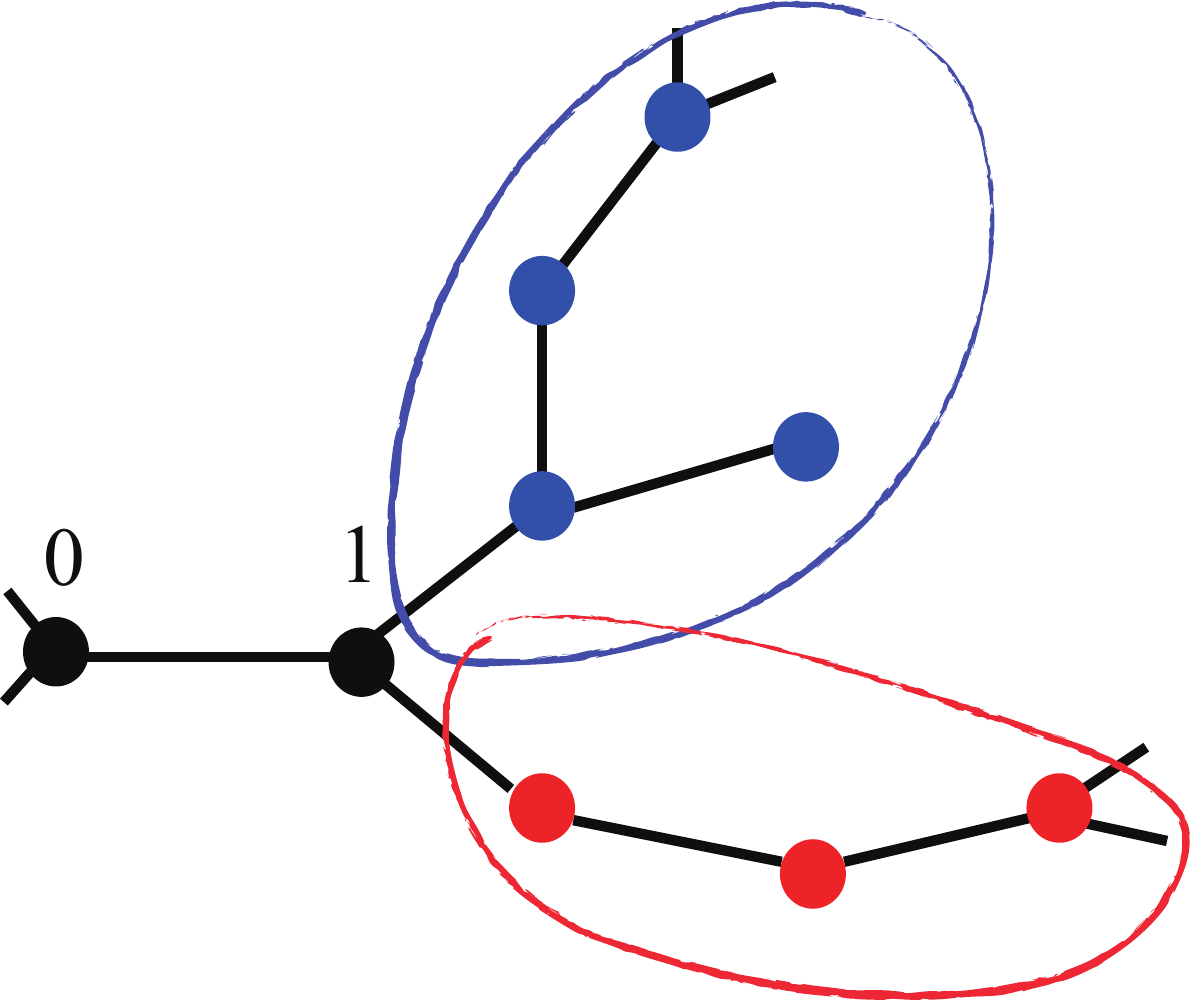}}
  \caption{(a) Sketch of a fragment of a Bethe lattice with $z=3$ with occupied (full circles) and unoccupied (empty circles) sites. A tracer on site 0 connects to one of the branches through site $1$. (b) A sketch of a cluster branch of size $s=8$. This branch can be further decomposed into two sub-branches of sizes $s_1=4$ (blue) and $s_2=3$ (red) connected by site $1$.}
  \label{fig:bethesketch}
\end{figure}

In this section, we extend the computation of Straley~\cite{straley1980ant}, 
to obtain the mean-square displacement for $p<p_\mathrm{c}$ on Bethe lattices of fixed general connectivity $z$
(Straley only considered the case $z=3$). Our main aim is to establish generically the behavior $\Delta^2(p)\sim A_\Delta(z)\ln|p-p_\mathrm{c}|$ and to study $A_\Delta(z)$, especially its large $z$ behavior, which is equivalent to the $d\rightarrow\infty$ limit in a hypercubic lattice. As $z\rightarrow\infty$, the logarithmic divergence of the mean-square displacement could indeed vanish and let weaker divergences or even proper caging emerge.

Consider a tracer initially on site $0$. By symmetry, we can examine any of the branches connected to $0$, hence, without loss of generality, we define the cluster as all occupied sites that are accessible to the tracer from site $0$ in one such branch.
Note that by convention the branch size does not include site $0$. (This convention, which differs from that of Ref.~\cite{straley1980ant}, simplifies the analysis for arbitrary $z$, whilst the alternate convention only simplifies the case $z=3$.) In addition to the branch size, $s$, another key quantity is the sum of chemical distance $Y = \sum_{i} d_{i0}$. On a finite-dimensional lattice, one would normally compute $\sum_{i} |{\bf r}_i-{\bf r}_0|^2$, but high-dimensional percolation clusters are effectively trees embedded on an hypercubic lattice.
Exploring the tree is thus equivalent to performing a random walk on that lattice, and because the chemical distance is the number of steps of the random walk, the distance on the hypercubic lattice is instead $d_{i0}\sim |{\bf r}_i-{\bf r}_0|^2$. (This identity has long been assumed in physics and is now rigorously proven to a large extent~\cite{aharony1984scaling,hara1990mean,arous2016scaling}). In summary, if the tracer belongs to a cluster of size $s$, the mean-square displacement for large $s$ is given by $Y/s$.  

Following Ref.~\cite{straley1980ant}, we then compute the average of this quantity over the cluster distribution. If site $1$ on the branch is not covered, then the branch size is $s=0$ and $Y = 0$. But if site $1$ is covered, then inserting site $0$ adds an additional step to each $(\mathbf{r}_i - \mathbf{r}_1) \rightarrow (\mathbf{r}_i - \mathbf{r}_0)$ for a tracer on site $0$ instead of site $1$. In other words, a branch with a tracer at site $0$ can be decomposed into $(z-1)$ subbranches that overlap at site $1$. The recurrence equations for $s$ and $Y$ thus read
\begin{align*} 
s &= s_1 + s_2 + ... + s_{z-1} + 1 \\
Y &= Y_1 + Y_2 + ... + Y_{z-1} + s,
\label{eqs:recur}
\end{align*}
and the probability of having a cluster with given $s$ and $Y$ is
\begin{equation}
\begin{aligned}
P(s, Y) = q\delta(Y,0)\delta(s,0) + p \sum_{s_1, ..., s_{z-1}; Y_1, ..., Y_{z-1}} \\
\delta(s, 1+\sum_{i=1}^{z-1}s_i) \delta(Y, s+\sum_{i=1}^{z-1}Y_i) \prod_{i=1}^{z-1}P(s_i, Y_i),
\end{aligned}
\end{equation}
where $q\equiv 1-p$ denotes the probability that a site not be covered. Taking the Laplace transform, we obtain
\begin{equation} \label{eq:laplaceP}
\hat{P}(x) = \sum_{s, Y} e^{-x s} P(s, Y) = q + p e^{-x} \hat{P}(x)^{z-1}
\end{equation}
and
\begin{equation}
\begin{aligned} \label{eq:laplacePY}
\hat{P_Y}(x) &= \sum_{s, Y} e^{-x s} Y P(s, Y) \\
&= -p \dv{(e^{-x} \hat{P}(x)^{z-1})}{x} + (z-1) p e^{-x} \hat{P_Y}(x) \hat{P}(x)^{z-2} \\
&= -\dv{\hat{P}(x)}{x} + (z-1) p e^{-x} \hat{P_Y}(x) \hat{P}(x)^{z-2}.
\end{aligned}
\end{equation}
Considering the whole cluster as $z$ independent branches joined at site $0$, we can then write the expected displacement
\begin{equation} \label{eq:intdispx}
\begin{aligned}
\Delta^2 &= \angles{\frac{\sum_{i=1}^z Y_i}{\sum_{i=1}^z s_i}} \\
&= \sum_{s,Y} \left( \frac{Y_1 + Y_2 + ... + Y_{z}}{s_1 + s_2 + ... + s_{z-1} + 1} \prod_{i=1}^{z} P_i(Y_i, s_i) \right) \\
&= z \int_0^\infty e^{-x} \hat{P_Y}(x) \hat{P}(x)^{z-1} \dd x.
\end{aligned}
\end{equation}

Note that from Eqs.~\eqref{eq:laplaceP} and~\eqref{eq:laplacePY}, we have
\begin{align*}
e^{-x} &= \frac{\hat{P} - q}{p \hat{P}^{z-1}} \\
\hat{P_Y} &= \dv{\hat{P}}{x} \left( \frac{\hat{P}(x)}{\hat{P}(x)(z-2) - q(z-1)}\right)
\end{align*}
Because $\hat{P}(0) = 1$ and $\lim_{x \rightarrow \infty}\hat{P}(x) = q$, we can change the integration variable from $x$ to $\hat{P}$ in  Eq.~\eqref{eq:intdispx} to obtain
\begin{equation} \label{eq:intdispP}
\begin{aligned}
\Delta^2 &= -\frac{z}{p} \int_{q}^{1} \frac{\hat{P}(\hat{P}-q)}{\hat{P}(z-2) - q(z-1)} \dd \hat{P} \\
&= -\frac{z(zq+z-2)}{2(z-2)^2} + \frac{q^2(z-1)z}{p(z-2)^3} \ln \frac{q p_\mathrm{c}}{p_\mathrm{c} - p},
\end{aligned}
\end{equation}
where for a Bethe lattice $p_\mathrm{c} = 1/(z-1)$. In the limit $p \rightarrow p_\mathrm{c}^-$, we finally have 
\begin{equation} \label{eq:YNtrend}
\begin{aligned}
\Delta^2 &\sim \frac{q^2(z-1)z}{p(z-2)^3} \ln [(p_\mathrm{c} - p)^{-1}] \\
&= A_\Delta (z)\ln [(p_\mathrm{c} - p)^{-1}],
\end{aligned}
\end{equation}
with $A_\Delta(z)=z/(z-2)$. For $z=3$, this expression reduces precisely to that of Ref.~\cite[Eq.~(2.10)]{straley1980ant}. Because in the limit $z \rightarrow \infty$, $A_\Delta(z) \rightarrow 1$,  we conclude that $\Delta^2(p\rightarrow p_\mathrm{c})$ diverges logarithmically in \emph{all} dimensions. 
 
\subsection{Random graph analysis} \label{sec:rgana}

As a validation of the $d\rightarrow \infty$ result, we separately consider confinement upon approaching percolation on a random graph, which directly evaluates the fully-connected limit. 
For convenience, we specifically consider Erd\"os-R\'enyi random graphs~\cite{erdos1960evolution}, which are obtained by considering $s\rightarrow \infty$ vertices and connecting each pair of vertices by an edge with probability $p/s$, where $p$ does not scale with $s$, i.e., $p \propto \mathcal{O}(1)$. The percolation threshold is then $p_\mathrm{c}=1$. 

In order to compute $\Delta^2$, we take site $0$ uniformly at random as the initial position of the tracer, and then let that tracer diffuse for an infinite amount of time. As on a Bethe lattice, we then have
\begin{equation}
\Delta^2=\angles{\sum_{i\in \mathbb{S}} \frac{d_{0i}}{s}},
\end{equation}
where the sum is over all sites that belong to the cluster, $\mathbb{S}$, to which site $0$ belongs, $s=|\mathbb{S}|$ is the cluster's size,
and the chemical distance $d_{0i}$ is here the smallest number of edges needed to go from site $0$ to site $i$. 
The factor of $1/s$ accounts for the equiprobability of each site in the infinite-time limit.
In the notation of Sect.~\ref{sec:bethe}, we have $Y=\sum_{i \in \mathbb{S}} d_{0i}$, which recovers $\Delta^2=\langle Y/s \rangle$, as above. The only distinction is that we here treat the whole cluster at once and do not identify a specific cluster branch. 

In order to compute $Y$, we study the probability distribution function $P(s,Y)$ using recurrence relations. Adding a new vertex to a graph with $s$ vertices merges into a single cluster all clusters that contain this vertex (recalling that all such clusters are finite because we are considering $p<p_\mathrm{c}$), and hence
\begin{align*}
s &= s_1+\cdots+s_k+1, \\
Y &= \sum_{l=1}^k \sum_{i\in \mathbb{S}_l} (d_{i,\mathrm{new}}+1), 
\end{align*}
where $s_1, s_2, ..., s_k$ are the sizes of the clusters to which the new vertex is attached. We can thus write that for the new site
\begin{equation} \label{eq:rgrecur}
\begin{aligned}
P_{s+1}(s,Y) = \rho_0 \delta(s,1) \delta(Y,0) + \sum_{k=1}^s \rho_k \sum_{s_1,\dots,s_k;Y_1,\dots,Y_k} \\
\delta(s,1+\sum_{i=1}^k s_i)\delta(Y,\sum_{i=1}^k Y_i+s_i)\prod_{i=1}^k P_s(s_i,Y_i) ,
\end{aligned}
\end{equation}
where $\rho_k=\frac{p^k}{k!}e^{-p}$ is the probability that a given site has exactly $k$ neighbors.

Introducing the generating function
\begin{equation*}
G(y,z)=\sum_{s,Y}e^{-yY-zs}P(s,Y),
\end{equation*}
we can rewrite Eq.~\eqref{eq:rgrecur} as
\begin{equation}\label{sc}
G(y,z)=e^{-z}\exp\left[p\left(G(y,z+y)-1\right) \right],
\end{equation}
because in the large $s$ limit the right- and left-hand sides of that distribution have the same limit. We then have
\begin{equation}
\Delta^2=\langle Y/s \rangle=-\int_0^\infty \dd z \left.\partial_y G(y,z)\right|_{y=0},
\end{equation}
where Eq.~\eqref{sc} gives
\begin{equation*}
\left.\partial_y G(y,z)\right|_{y=0}=\frac{pG(0,z)\partial_z G(0,z)}{1-pG(0,z)}.
\end{equation*}
We finally obtain
\begin{equation}
\begin{aligned}
\Delta^2 &=-\int_0^\infty \dd z \frac{pG(0,z)\partial_z G(0,z)}{1-pG(0,z)} \\
&=\int_0^1 \frac{pG}{1-pG}\dd G=-1-\frac 1 p \ln(1-p),
\end{aligned}
\end{equation}
which logarithmically diverges when $p\rightarrow p_\mathrm{c}=1$ with prefactor $A_\Delta(\infty)=1$. This result is therefore fully consistent with the limit of infinite connectivity, $z\rightarrow\infty$, for Bethe lattices considered in Sect.~\ref{sec:bethe}, and confirms that a logarithmic divergence of $\Delta^2(p\rightarrow\infty)$ is expected in all $d\geq d_\mathrm{u}$ with a monotonically decreasing, but non-vanishing prefactor $A_\Delta(z)$.

\section{Numerical Simulations of a Hypercubic Lattice}
\label{sec:simulations}

In order to validate the above scaling predictions numerically, we separately consider the scaling of the mean-square displacement on systems around the percolation threshold and evaluate the dynamical exponent directly at the percolation threshold. This section details the two computational schemes employed.

\subsection{Generating clusters}
Below the percolation threshold the mean-square displacement of a random walk eventually approaches a finite-height plateau. Because the finite state of the tracer decorrelates from its initial position in the infinite-time limit, we then have
\begin{equation} \label{eq:msdlimdef}
  \Delta^2(p)
  = \sum_{\{\mathbb{S}\}} \frac{P(\mathbb{S},p)}{s^2} \sum_{i,j \in \mathbb{S}} R_{ij}^2,
\end{equation}
where $\mathbb{S}$ refers to a given finite cluster of size $s=|\mathbb{S}|$, and $P(\mathbb{S},p)$ is the probability that a tracer falls within this cluster at covering fraction $p$. Note that the second summation is over all pairs of sites within $\mathbb{S}$.

In order to minimize the contribution of finite-size corrections, we implement the Leath algorithm~\cite{leath1976clustersize,mertens2018percolation}, which grows a cluster from the origin outward without storing the whole hypercubic lattice that embeds it. Under this sampling scheme, Eq.~\eqref{eq:msdlimdef} becomes
\begin{equation} \label{eq:msdlimleath}
\begin{aligned}
  \Delta^2(p)
  &= \left[\sum_{\{\mathbb{S}\}}  \frac{1}{ s^2 } \sum_{i,j \in \mathbb{S}} R_{ij}^2 \right] \\
  &= \left[ \sum_{\{\mathbb{S}\}}  2 ( \angles{R_i^2} -  \angles{R_i}^2) \right] ,
\end{aligned}
\end{equation}
where the inner average $\langle\ldots\rangle$ is over the different cluster sites and the outer average $[\ldots]$ is over the clusters generated.
Because periodic boundary conditions are not employed by the Leath algorithm, a percolating cluster cannot be generated for any $p<p_\mathrm{c}$. Finite-size effects are then of a different nature; they arise when the cluster volume becomes larger than the available memory of the computer (In practice we here use 60 GB). This weaker size constraint allows us to push computations by at least one order of magnitude in $|\epsilon|$ compared to the direct generation of a periodic hypercubic lattice.

Note that in our implementation, the coordinates of the visited sites, which include both the cluster and the neighboring sites (the cluster perimeter), are stored in a tree-based set. While hash tables are generally used to index sites in the context of percolation~\cite{grassberger2003critical,mertens2018percolation}, an efficient hash function that would limit hashing collisions even for large clusters is challenging to design. Because hashing collisions increase the search complexity by the maximal size, $\mathcal{O}(n)$, of the associated linked lists, their computational cost can grow quickly for large clusters. (Although rare, large clusters contribute most to $\Delta^2$ as $\epsilon \rightarrow 0^-$.) Tree-based sets, by contrast, cap the complexity at $\mathcal{O}(\log N)$, irrespective of the implementation, and were found to be more robustly efficient in the size regime studied here.

\subsection{Dynamics}

In order to probe the time evolution of the tracer, one may implement a dynamical equivalent of the Leath algorithm. Specifically, a tracer (blind ant) is first placed at the origin and then performs a random walk, attempting to jump over an edge to one of the neighboring sites with equal probability $p_\mathrm{b}=1/2d$. (Other tracer dynamics are possible, but the critical behavior is unaffected by this choice.) If the attempted site has never been visited, occupancy of that site is assigned with probability $p$, and the coordinates and occupancy of that site are stored as key-value pairs in a map. If the site has been previously deemed occupied (or vacant), then the tracer position is updated (or not) and time is incremented. 

While straightforward to implement, this brute-force method encounters a couple of difficulties at and around $p_\mathrm{c}$. First, the time required to approach the asymptotic scaling can be long, especially near the upper critical dimension. For instance, performing a random walk on a single cluster up to $t=10^{10}$ steps takes minutes, and is barely sufficient in $d=6$. Second, Eq.~\eqref{eq:pl} suggests that while both clusters with $s\lesssim s^*$ and $s>s^*$ contribute equally to $D^2(t)$, the latter are rarely generated by the Leath algorithm. In fact, the probability of generating clusters of size $s>s^*$ scales as $s^{2-\tau}$.
Assuming that a fixed number of clusters of size $s > s^*$ is needed to obtain reliable estimates at times $t \sim (s^*)^{d_\mathrm{w}/d_\mathrm{f}}$,
then the total number of samples, $N_\mathrm{sample}$, should grow with $t$ as
\begin{equation} \label{eq:dynvar}
N_\mathrm{sample} \sim 1/(s^*)^{2-\tau} \sim t^{\frac{d_\mathrm{f}}{d_\mathrm{w}}(\tau-2)},
\end{equation}
where the net power-law exponent is about $0.12, 0.20, 0.26$ and $0.33$ for $d=3,4,5$ and $d \rightarrow 6^-$, respectively.
In other words, in order to keep the accuracy of $D^2(t)$ constant while increasing $t$ by an order of magnitude, one has to generate 1.3, 1.5, 1.8 and 2.1 times more clusters, respectively. The scale of the numerical challenge thus increases with $d \le d_\mathrm{u}$. (For $d \ge d_\mathrm{u}$,  the computational difficulty remains roughly constant because the critical exponents remain unchanged.)

In order to mediate these issues, we devise a scheme to compute the dynamical contribution of small and intermediate size clusters without explicitly simulating the random walk. Our approach relies on approximating the mean-square displacement of clusters of size $s < s^*$ by its infinite-time limit plus corrections (see Eq.~\eqref{eq:pl}). One then needs to determine an appropriate $s^*$ (not only its scaling) at a given time $t$, or, equivalently, for a given cluster determine the relaxation time, $t^*$, such that for $t > t^*$ the mean-square displacement can be similarly approximated.

More specifically, consider a cluster $\mathbb{S}$ with sites $i = 0, 1, ..., s-1$, and denote $P(i, j, t)$ the probability that a tracer at site $i$ travels to site $j$ in $t$ steps. For instance, for $t=1$, we have
\begin{equation}
P(i, j, t=1) = \begin{cases}
\frac{1}{2d}, & j \in \partial{i}, \\
0, & j \not\in  \partial{i} \text{ and } i \neq j, \\
1 - \frac{|\partial i|}{2d}, & i=j,
\end{cases}
\end{equation}
where $\partial i$ denotes sites neighboring site $i$. Obviously, we have $P(i,j,1) = P(j,i,1)$, and for $t>1$ 
\begin{equation}
P(i, j, t) = \sum_{k \in \mathbb{S}} P(i, k, t-1) P(k, j, 1).
\end{equation}
For convenience, we define the transfer matrix of diffusion with entries $\mathbf{P}_{ij} = P(i,j,1)$, and hence $P(i,j,t) = \mathbf{P}^t_{ij}$. The mean-square displacement on $\mathbb{S}$ is then
\begin{equation} \label{eq:tmat}
D^2(t) = \frac{1}{s} \Tr\left( \mathbf{P}^t \mathbf{R^2} \right) =  \frac{1}{s} \Tr\left( \mathbf{\Lambda}^t \mathbf{Q}^\mathrm{T} \mathbf{R^2} \mathbf{Q} \right),
\end{equation}
where the entry $\mathbf{R^2}_{ij}$ is the square distance between sites $i$ and $j$. Because $\mathbf{P}$ is real and symmetric, its eigendecomposition gives orthogonal eigenvectors, such that $\mathbf{P} =  \mathbf{Q} \mathbf{\Lambda}  \mathbf{Q}^\mathrm{T}$.
One can then straightforwardly obtain $D^2(t)$ of a given cluster at arbitrary $t$ by Eq.~\eqref{eq:tmat}. 

From this scheme, it is also possible to determine the asymptotic dynamics on a finite cluster with arbitrary accuracy.
Because $D^2(t \rightarrow \infty) = \mathrm{cnst}$, the leading eigenvalue of $\mathbf{P}$ is $\Lambda_0 = 1$ and the corresponding eigenvector $\bm{q}_0$ has identical elements $\bm{q}_i =1/s (\forall i=0 \text{ to } s-1$). For a cluster $\mathbb{S}$, we can then consider the $i$-th dynamical relaxation time in relation to the $i$-th subleading eigenvalue, $\Lambda_i$, 
\begin{equation}
t^*_i(\mathbb{S}) = -1/\ln \frac{\Lambda_{i}}{\Lambda_{0}} = -1/\ln \Lambda_{i},
\end{equation}
and thence
\begin{equation} \label{eq:relaxtime}
D^2(t) = \Delta^2 + \sum_{i=1}^{s-1} \bm{q}^\mathrm{T}_i \mathbf{R^2}  \bm{q}_i e^{-t/t^*_i} = \Delta^2 - \sum_{i=1}^{s-1}c_i e^{-t/t^*_i}.
\end{equation}
with prefactor $c_i = -\bm{q}^\mathrm{T}_i \mathbf{R^2} \bm{q}_i$. Because $|c_i|$ tends to decrease with $i$, $t^*_i$ provides a rough upper bound on the time at which the pre-asymptotic corrections due by all $c_j$ with $j \ge i$ are significant. In other words, if one uses the first $i$ leading eigenvalues and eigenvectors to approximate $D^2(t)$, the result is robust when
\begin{align}
\frac{c_i}{\Delta^2} e^{-t/t^*_i} &\ll 1, \nonumber \\
\Rightarrow t > t^*_i &= -1/\ln \Lambda_i \approx \frac{1}{1-\Lambda_i}.
\end{align}
The pre-asymptotic behavior of $D^2(t)$ can thus be evaluated directly at short times, and computed by Eq.~\eqref{eq:relaxtime} for $t > t^*_i$. 

In particular, the mean-square displacement of a tracer on a single cluster of size $s^* \sim (t^*_1)^{d_\mathrm{f}/d_\mathrm{w}}$ approaches a constant for $t>t^*_1=-1/\ln \Lambda_{1}$. This relation suggests that the distance between the second leading eigenvalue and unity, $1-\Lambda_1 \sim s^{-d_\mathrm{w}/d_\mathrm{f}}$, sets the maximal cluster size that can be treated by the transfer matrix approach for a finite machine precision. Here, we apply the transfer matrix approach only for the clusters smaller than the threshold cluster size $s_\mathrm{t}=5000$. The machine error under double precision is then of order $10^{-10}$ times smaller than $1-\Lambda_1$.

Our detailed implementation runs as follows. We first apply the Leath algorithm to generate a cluster. If the algorithm stops with $s < s_\mathrm{t}$, then the transfer matrix approach is used to compute the long-time dynamics and a direct evaluation of the dynamics is run for $1 \le t < t^*_i$, where $i\leq 19$, i.e., up to the $20$-th leading eigenvalue. If the Leath algorithm returns a cluster with a boundary that is not closed, then we place a tracer at the origin and simulate the random walk by brute force. Because the tracer can then reach undetermined sites, the cluster keeps growing along the walk. 
Because these brute-force simulations are only performed for clusters with $s \ge s_\mathrm{t}$, which are rare, using the transfer matrix approach accelerates the computation by orders of magnitude, even without carefully optimizing $s_\mathrm{t}$ and the eigensolver.

\section{Results and Discussion}
In this section, we evaluate the caging and subdiffusive behavior of tracers on hypercubic lattices by the simulation methods described in Sect.~\ref{sec:simulations}, and compare the results with the theoretical predictions presented in Sect.~\ref{sec:theory}.

\label{sec:results}
\subsection{Cluster Size Distribution}
\begin{figure}
  \subfloat{\subfigimg[width=0.48\textwidth]{a}{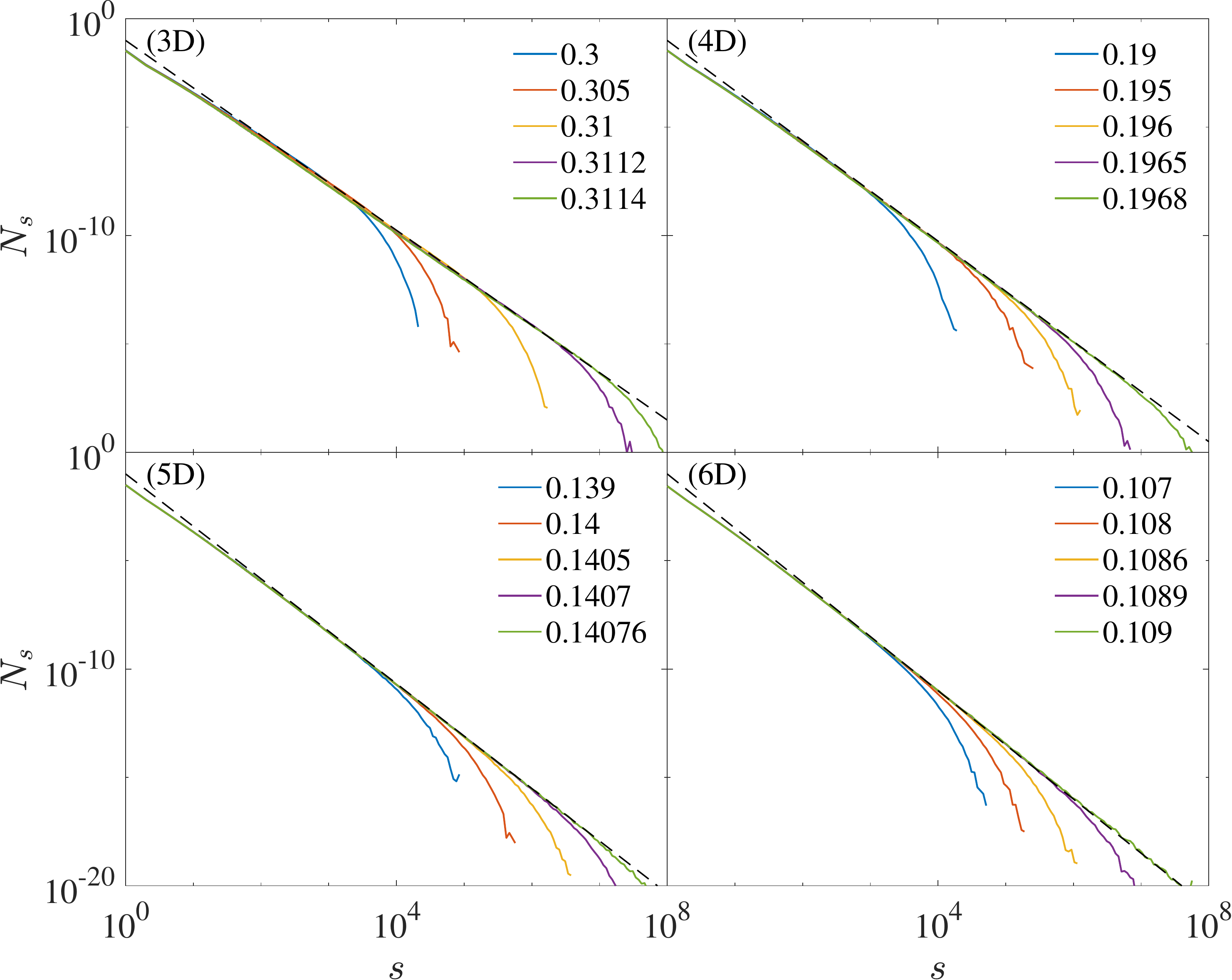}}\\
  \subfloat{\subfigimg[width=0.48\textwidth]{b}{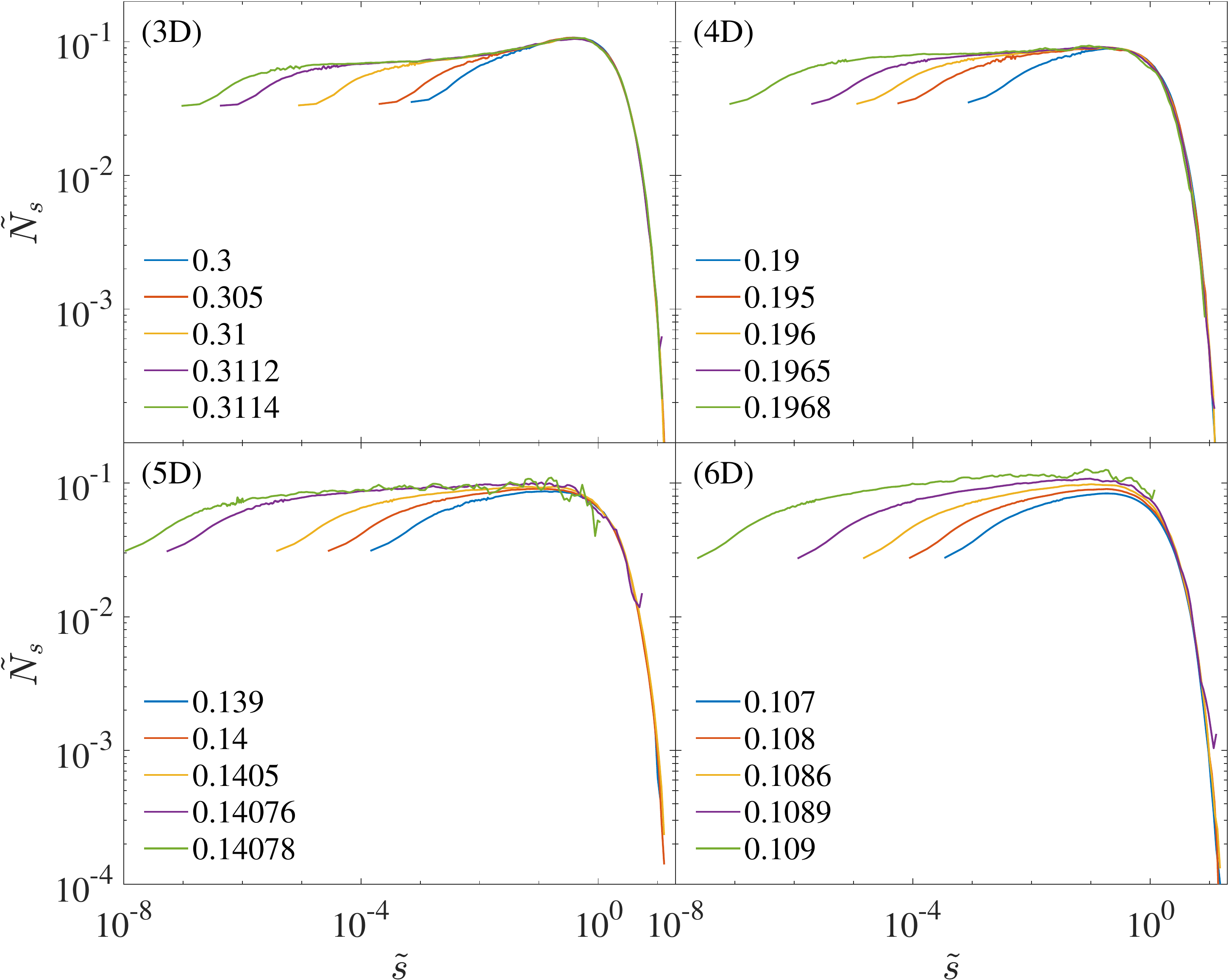}}
  \caption{(a) Cluster size distributions, $N_s(s, p)$, for different $p<p_\mathrm{c}$ in $d=3,4,5$, 6 (as given in panel labels). The Fisher power-law scaling (dashed line) describes well the large-cluster regime. (b) The rescaled cluster size distribution $\tilde{N_s}(\tilde{s}, p)$, shows a very good collapse for the exponential cutoff of the cluster size distribution in all dimensions, but more sizable pre-asymptotic effects for the Fisher tail can be observed as $d$ increases.}
    \label{fig:cludist}
\end{figure}

We first consider the distribution of cluster sizes, $N_s$, generated by the Leath algorithm near the percolation threshold. Figure~\ref{fig:cludist}(a) suggests that for $\epsilon\rightarrow0$, the intermediate decay of $N_s$ can be fitted to a power law; at larger sizes, $N_s$ systematically deviates from this scaling, but the closer $p$ is to $p_\mathrm{c}$, the more extended the power-law scaling regime. The specific scaling prediction for the cluster distribution is that
\begin{equation} \label{eq:fisher}
 N_s(p) = s^{-\tau} (f_0(\tilde{s}) + s^{-\Omega} f_1(\tilde{s}) + ...),
\end{equation}
for the rescaled cluster size $\tilde{s} = |\epsilon|^{1/\sigma}s$. This form suggests that the largest likely cluster scales as $s_\mathrm{max} \sim |\epsilon|^{-1/\sigma}$, before the cluster size distribution deviates from the Fisher power law, and that results can be asymptotically collapsed by considering the rescaled function $\tilde{N}_s = s^\tau N_s$ vs $\tilde{s}$. 

Figure~\ref{fig:cludist}(b) shows that for the reference exponents $\tau$ and $\sigma$ (Table~\ref{tab:refdata}),  all cluster size distributions asymptotically collapse onto a single master function. Because previous numerical results for $N_s$ either considered a finite periodic box~\cite{christensen2005complexity} or truncated the cluster size distribution before $s_\mathrm{max}$ could be reached~\cite{mertens2018percolation}, this master function had not before been seen all at once. The Leath algorithm here enables us to grow directly clusters up to $10^8$ sites, hence the master curve clearly displays the two regimes: a flat $\tau$-dominated regime and a sharply decaying $\sigma$-dominated regime, with a crossover around $s_\mathrm{max}$. 

In addition to validating our implementation of the Leath algorithm, this analysis allows us to validate the reported values of the relevant critical exponents. While the most recent estimates $\tau$ are seemingly very accurate~\cite{mertens2018percolation}, those for $\sigma$ are not all consistent~\cite{ballesteros1999scaling,gracey2015four}. In $d=3$, in particular, the simulation estimate for $\sigma$~\cite{ballesteros1999scaling} leads to a better rescaling than the best renormalization group (RG) prediction~\cite{gracey2015four}, especially in the $\sigma$-dominated regime. Because increasing $d$ closer to $d_\mathrm{u}$ makes the RG predictions increasingly accurate, for the rest of our analysis, we use $\sigma$ determined by simulations in $d=3$ and by the RG treatment in $d=4$ and 5. 

\subsection{Critical Caging Regime}

\begin{figure*}
  \subfloat{\subfigimg[width=0.48\textwidth]{a}{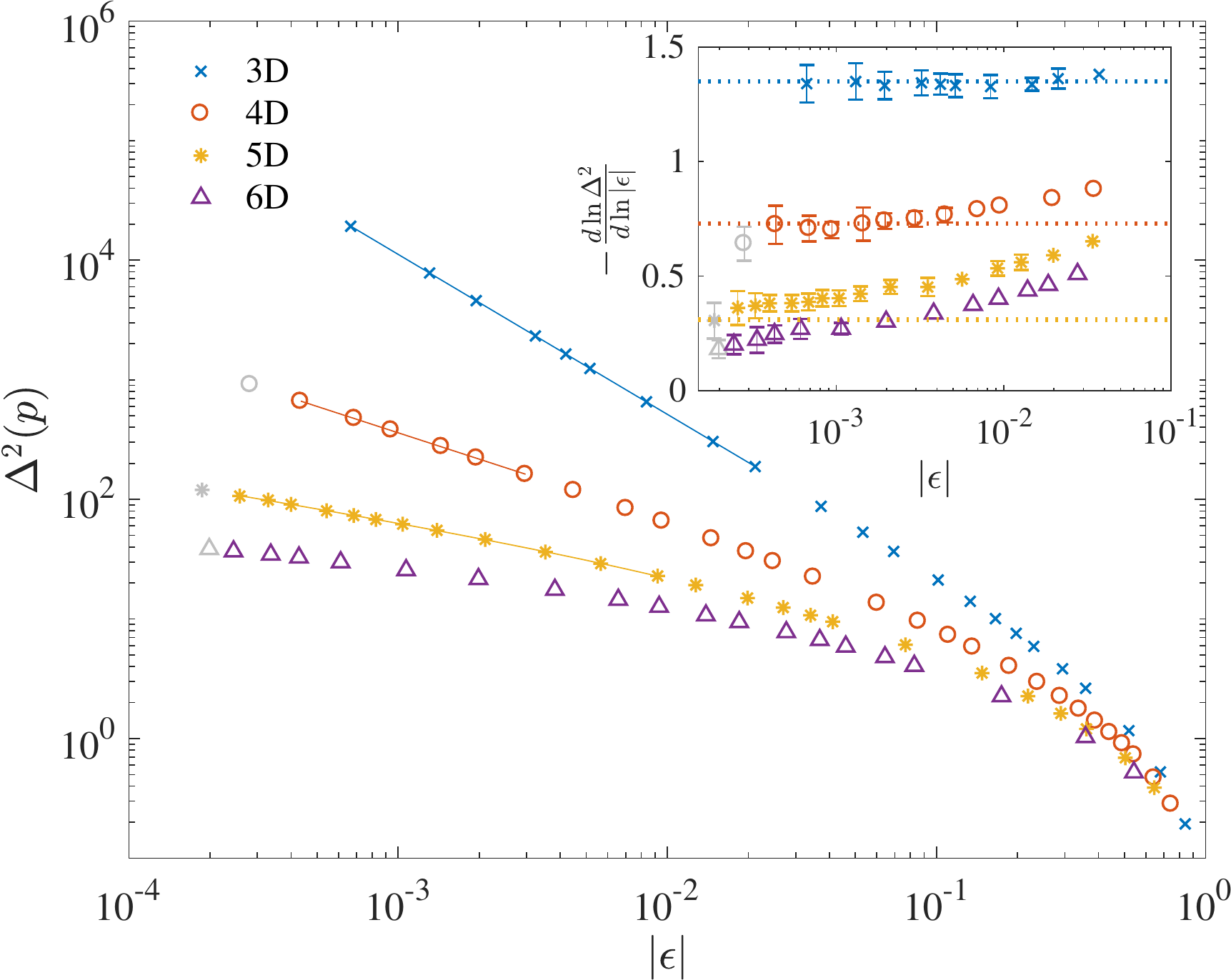}}
   \subfloat{\subfigimg[width=0.48\textwidth]{b}{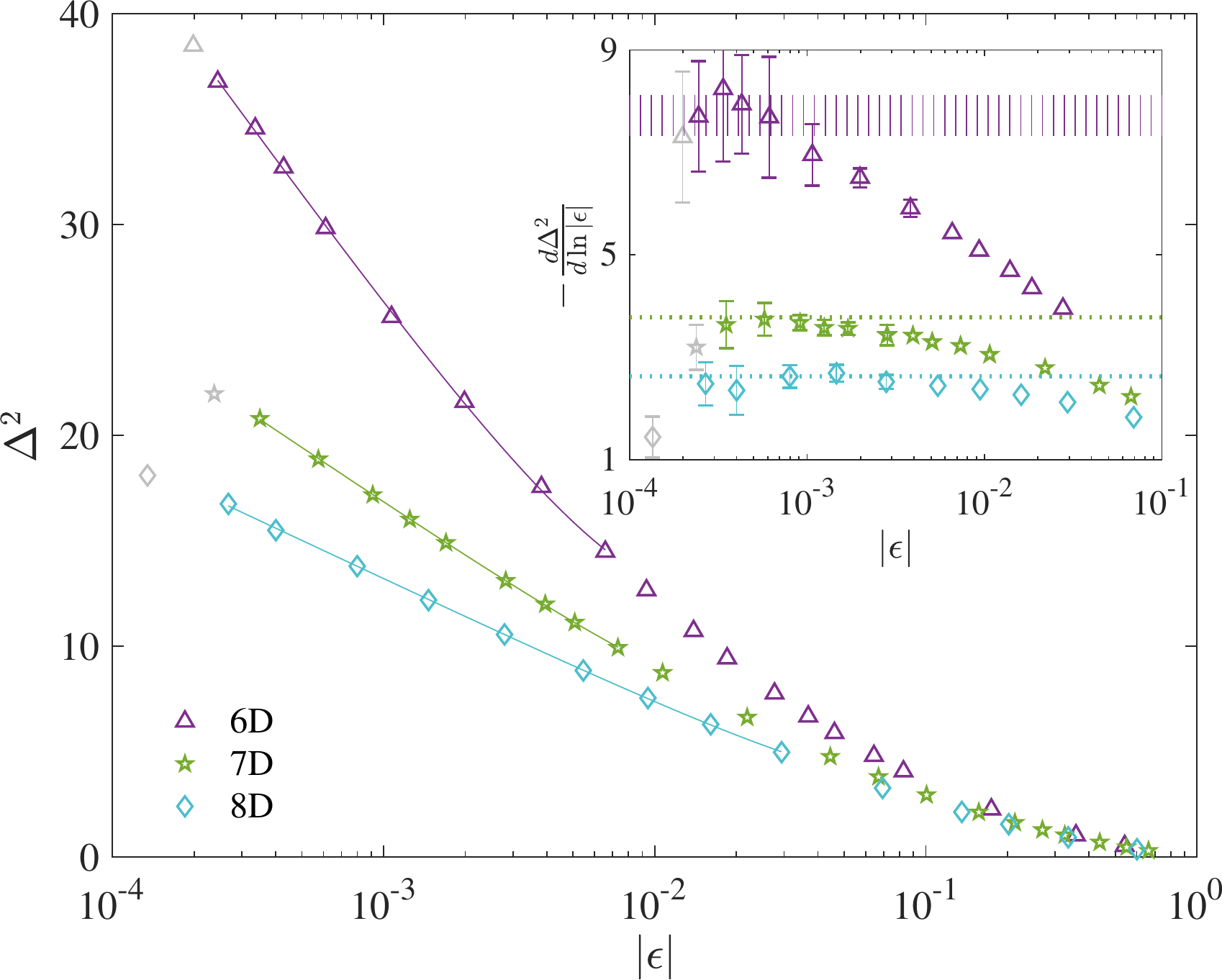}}
  \caption{$\Delta^2$ on (a) log-log scale for $d=3$ to 6 and (b) on a lin-log scale for $d=6$ to 8. Red data points are used for estimating $\mu_-$ in Table~\ref{tab:msdlim}. Gray points denote conditions, under which at least $10^{-3}\%$ of the clusters reached the maximal memory size, and are therefore numerically suspicious. These points are provided for context alone, they are not used in the analysis. (inset) (a) Estimates of $\mu_-$ obtained from the local slope of $\Delta^2$ by Eq.~\eqref{eq:localslope}, and (b) estimates for the prefactors obtained from the finite differentiation. Error bars denote $95\%$ confidence intervals. Dotted lines are (a) the scaling prediction given by Eq.~\eqref{eq:muminus_2} and (b) the fitting results of the main plot. The error bar on the estimate for $d=6$ is denoted by vertical strips.}
    \label{fig:msdlim}
\end{figure*}

Using the clusters generated by the Leath algorithm, we next compute $\Delta^2(p)$ for a tracer using Eq.~\eqref{eq:msdlimleath} (Fig.~\ref{fig:msdlim}). 
Directly evaluating the critical exponent $\mu_-$ is challenging because of the growing variance of the cluster size distribution as $\epsilon\rightarrow0$, and because of the sizable pre-asymptotic corrections. In order to assess the convergence of our results and identify the pre-asymptotic regime, we use the estimate
\begin{equation} 
\mu_- = -\frac{d \ln(\Delta^2)}{d \ln |\epsilon|} = \frac{(p_\mathrm{c}-p)}{\Delta^2} \frac{d \Delta^2(p)}{d p},
\end{equation}
which becomes exact in the limit $\epsilon\rightarrow0$.
Clusters being grown by random addition, the probability that a site belongs to any generated cluster $i \in \mathbb{S}$, with cluster size $s$ and perimeter $b$, is uniformly distributed. The weight of this cluster upon changing $p$ to $p'$ is then~\cite{grassberger2003critical}
\begin{equation} \label{eq:weightcluster}
\frac{w(p')}{w(p)} = \left( \frac{p'}{p} \right)^s \left( \frac{1-p'}{1-p}\right)^b,
\end{equation}
and $\Delta^2(p') = \angles{D^2}_{w(p')}$ is the weighted average of limiting the mean-square displacement, $D^2$, for every cluster $i$ sampled,
\begin{equation} \label{eq:msdsampleexpand}
\Delta^2(p') = \frac{\sum_{i=1}^N w_i D^2_i}{\sum_{i=1}^N w_i},
\end{equation}
where $N$ is the number of clusters sampled. Because $D^2$ for one single cluster does not depend on $p$, we can calculate explicitly the derivative of Eq.~\eqref{eq:weightcluster},
\begin{equation}
\left.\frac{d w_i}{d p'}\right|_{p} = \frac{s}{p} - \frac{b}{1-p},
\end{equation}
and combine the result with Eq.~\eqref{eq:msdsampleexpand} to obtain
\begin{equation} \label{eq:localslope}
\begin{aligned}
\frac{d \Delta^2}{d p} = \frac{1}{N^2} \Big( &\sum_{i \in \mathbb{S} } (\frac{s}{p} - \frac{b}{1-p}) D^2_i \\
 - &\sum_{i \in \mathbb{S} } (\frac{s}{p} - \frac{b}{1-p}) \sum_iD^2_i \Big).
\end{aligned}
\end{equation}
It is therefore possible to obtain an estimate $\mu_-$ by generating a sufficiently large number of clusters and by then calculating the local slope for a given fixed $p$ (Fig.~\ref{fig:msdlim}, inset). As expected, the numerical results agree with those obtained by scaling relations in $d=3$ and $4$ at small enough $\epsilon$. For these two cases, we can directly read off the proper fitting range and the numerical estimates of $\mu_-$ (Table~\ref{tab:msdlim}). In $d=5$, however, small systematic deviations persist within the whole numerically accessible regime. Such a slow convergence to the asymptotic scaling is characteristic of approaching $d_\mathrm{u}$. In this case, we therefore employ a  form with correction terms, 
\begin{equation}
\ln \Delta^2 =  -\mu_- \ln |\epsilon| + C + B/\ln |\epsilon|
\end{equation}
to fit the pre-asymptotic estimates. Because the resulting estimate of $\mu_-$ is then to some degree sensitive to the fitting range, we choose that range, such that $\mu_-$ is minimal. The resulting critical exponent is then consistent with the scaling relation prediction for $d=5$ (Table~\ref{tab:msdlim}).

\begin{table}
\caption{Caging critical exponent $\mu_-$ obtained as described in the text. Scaling predictions use reference exponents in Table~\ref{tab:refdata}.}
\begin{tabular}{cccc}
\hline \hline
$d$ & This work & Eq.~\eqref{eq:muminus} & Eq.~\eqref{eq:muminus_2} \\
\hline
3 & 1.3377(15) &  1.35(2) & 1.335(2) \\
4 & 0.73(1) & 0.73(2) & 0.723 \\
5 & 0.31(2) & 0.310(6) & 0.301 \\
\hline
\end{tabular}
 \label{tab:msdlim}
\end{table}

For $d \ge d_\mathrm{u}$, scaling relations predict that $\mu_- = 0$, and the inset in Fig.~\ref{fig:msdlim} indeed indicates that no finite $\mu_-$ value is reached over the numerically accessible regime in $d=6$.   For $d=6$, 7 and 8, we instead consider the logarithmic scaling prediction, $\Delta^2\approx -A_\Delta(d) \ln |\epsilon|$, with a dimensional-dependent prefactor $A_\Delta(d)$. Figure~\ref{fig:msdlim}(b) repeats the above analysis for this new scaling form.
For the critically marginal $d=6$, the growth of $\Delta^2$ seems to be accompanied by a fairly wide pre-asymptotic regime under the local slope analysis. It is also likely that $\Delta^2$ actually scales with a different form, such as that given in Eq.~\eqref{eq:6dscaling}. We come back to this hypothesis below. For now, we treat the $d \ge 6$ results on the same footing by using the fitting form
\begin{equation}
\Delta^2 = -A_\Delta \ln |\epsilon| + B_\Delta + C_\Delta \epsilon,
\end{equation}
which gives $A_\Delta(6)=7.7(4)$, $A_\Delta(7)=3.79(9)$ and $A_\Delta(8)=2.62(6)$. Unsurprisingly, these values differ from the Bethe lattice predictions for $d=2z$, which would be roughly 1.2, 1.17, and 1.14, respectively. The prefactors do, however, monotonically decrease with increasing $d$ and remain above unity $1$, which is the expected $d\rightarrow\infty$ prefactor for both Bethe and hypercubic lattices.

\subsection{Subdiffusive Scaling}

\begin{figure*}
  \subfloat{\subfigimg[width=0.48\textwidth]{a}{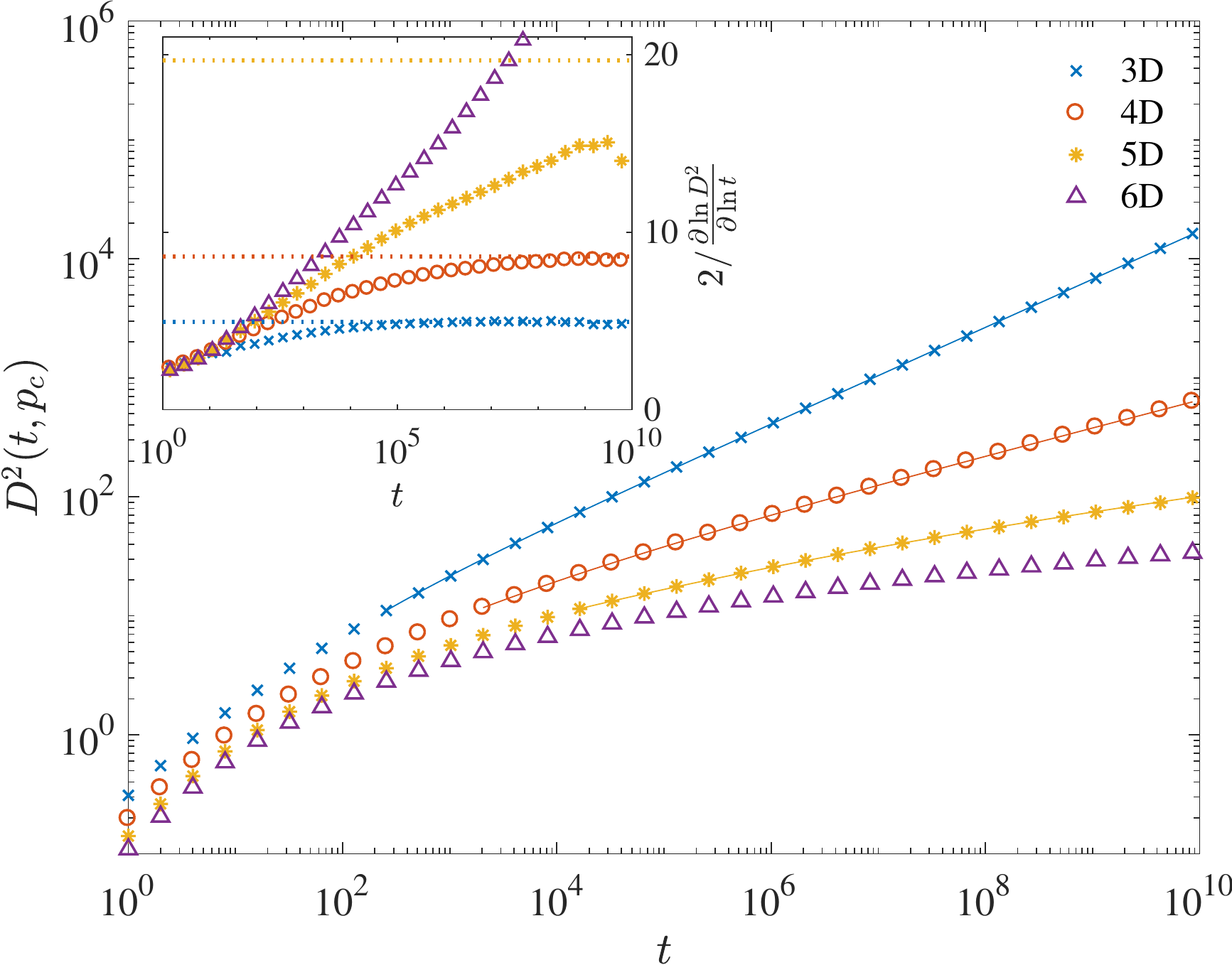}}
  \subfloat{\subfigimg[width=0.48\textwidth]{b}{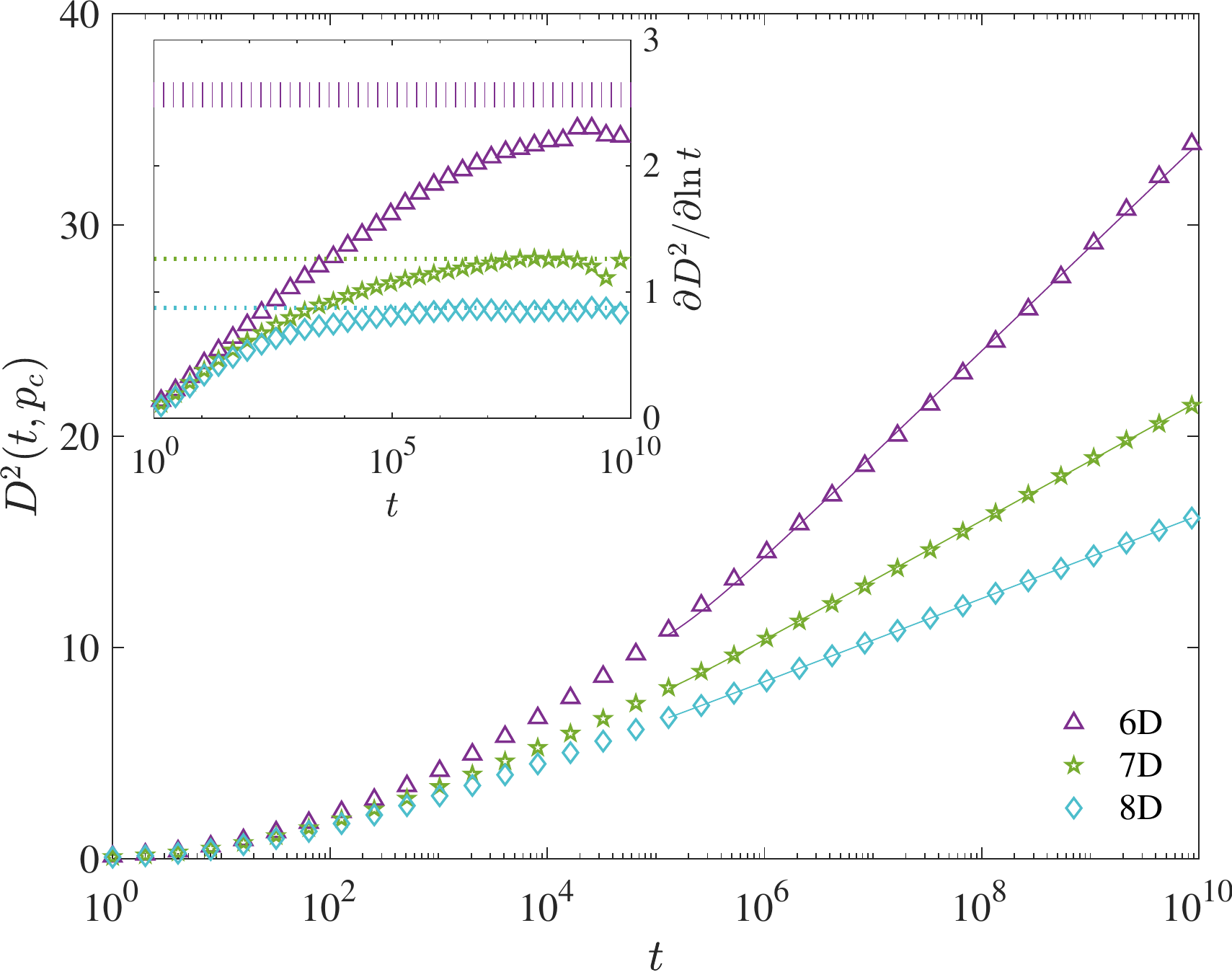}}
  \caption{$D^2(p_\mathrm{c})$ on (a) log-log scale for $d=3$ to 6, and (b) on a lin-log scale for $d=6$ to 8. Fitted results for $d'_\mathrm{w}$ are given in Table~\ref{tab:msdpp}. (inset) Pre-asymptotic corrections captured by the evolution of the local slope, i.e., (a) $2 / (\partial \ln D^2 / \partial \ln t)$ and (b) $(\partial D^2 / \partial \ln t)$. Dotted lines are (a) the scaling predictions given by Eq.~\eqref{eq:dwprime}, and (b) the predicted prefactor relationship, $A_D(d) = A_\Delta(d)/3$, where $A_\Delta(d)$ is obtained from Fig.~\ref{fig:msdlim}(b). Note that quantitative agreement for the prefactors is observed in $d=7$ and 8, but that in $d=6$ discrepancies are observed, despite the large error bar (See Fig.~\ref{fig:msdlim}(b))}
    \label{fig:msdpp}
\end{figure*}

Thanks to the availability of very precise estimates of the percolation threshold~\cite{mertens2018percolation}, $d'_\mathrm{w}$ can be obtained by running  dynamical simulations essentially \emph{at} $p_\mathrm{c}$. The growth of the mean-square displacement with time is shown in Fig.~\ref{fig:msdpp}. In order to estimate the pre-asymptotic scaling for $d < d_\mathrm{u}$, we consider the empirical form
\begin{equation} \label{eq:msdfit}
  D^2(t, p_\mathrm{c}) = C t^{2/{d'_\mathrm{w}}} + B + o(t^{-1}).
\end{equation}
where $B$ and $C$ are fitting constants. Note that in practice, the fit is done on a logarithmic scale, that is
\begin{equation} \label{eq:msdfitprac}
  \ln D^2(t, p_\mathrm{c}) \approx \frac{2}{d'_\mathrm{w}} \ln t + \ln C + \frac{B}{C} t^{-2/{d'_\mathrm{w}}}.
\end{equation}
We use $D^2 > 10$ as fitting range, which appears to fall near the end of the pre-asymptotic corrections that are not captured by this form.

\begin{table}
\caption{Subdiffusion exponent $d'_\mathrm{w}$ obtained as described in the text along with earlier numerical estimates. Scaling predictions use reference exponents in Table~\ref{tab:refdata}.}
\begin{tabular}{cccccc}
\hline \hline
$d$ & This work
& Eq.~\eqref{eq:dwprime} & Ref.~\cite{lee2003correction}\\
\hline
3 & 4.94(1) 
& 4.96(8) & 5.04(1)\\
4 & 8.64(4) 
& 8.6(2) & 8.37(1)\\
5 & 20(3) 
& 19.7(13) & -\\
\hline
\end{tabular}
 \label{tab:msdpp}
\end{table}

In order to identify the pre-asymptotic regime, we also consider the evolution of the local slope, $2 / (\partial \ln D^2 / \partial \ln t)$ with $\ln t$. Here again, for $d=3$ and $4$ the numerical estimates of $d'_\mathrm{w}$ agree with the scaling predictions given by Eq.~\eqref{eq:dwprime}. Note that our implementation of the transfer matrix scheme allows us to average over many more clusters and to run longer trajectories than was previously possible, which likely explain the discrepancy with older numerical estimates~\cite{lee2003correction}.
For $d=5$, however, the computationally accessible regime remains nonetheless somewhat distant from the asymptotic scaling.
Although no precise numerical estimate is thus available, Eq.~\eqref{eq:msdfitprac} plausibly gives $d_\mathrm{w} \approx 20$, which is consistent with the scaling prediction.

\begin{figure*}
  \subfloat{\subfigimg[width=0.98\textwidth]{a}{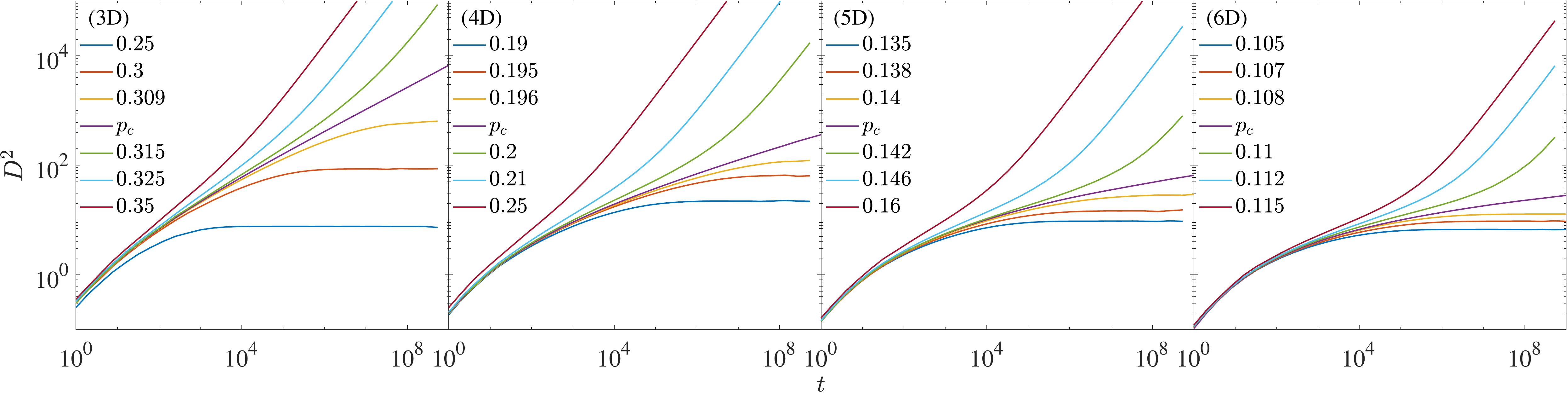}} \\
  \subfloat{\subfigimg[trim={-5 0 0 0},clip,width=0.98\textwidth]{b}{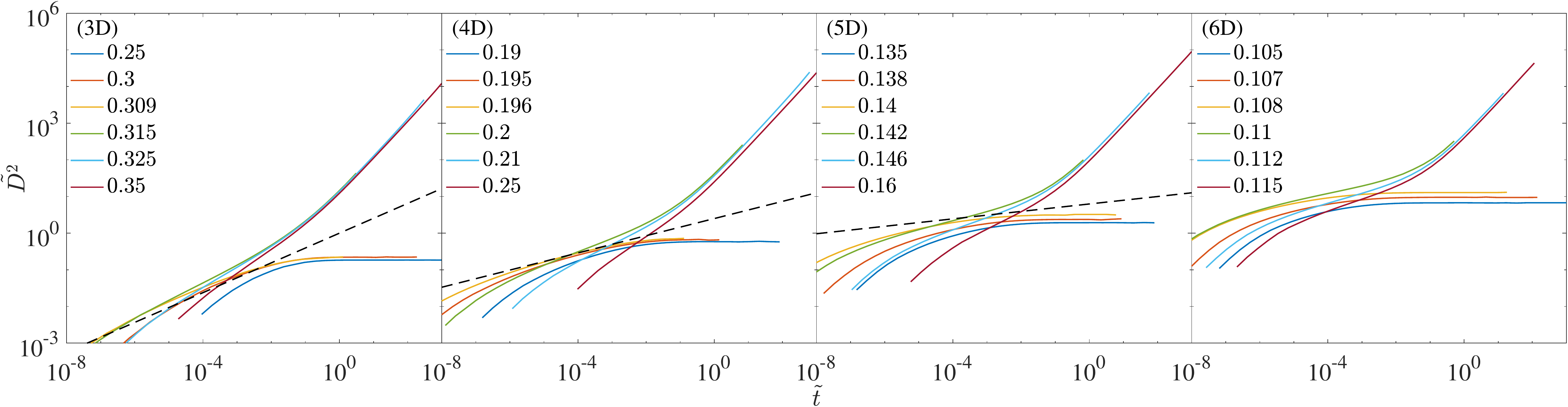}} \\
  \caption{(a) Original and (b) rescaled time and MSD for $d=3$, $4$, $5$, and $6$. The dashed-line in (b) denotes the subdiffusive scaling of $\tilde{D} \sim \tilde{t}^{2/d'_\mathrm{w}}$. }
  \label{fig:msddyn}
\end{figure*}

From the scaling analysis of Sect.~\ref{sec:theory}, we expect $\Delta^2(t)$ for $d \ge d_\mathrm{u}$ to grow logarithmically with time. Figure~\ref{fig:msddyn} explicitly tests this prediction, and we implement the fitting form
\begin{equation}
  D^2(t, p_\mathrm{c}) = A_D \ln t + B_D + C_D / t
\end{equation}
to approximate the prefactor $A_D$ of the logarithmic growth.
The numerical results supports the logarithm scaling, giving $A_D=2.14(4), 1.230(8)$ and $0.855(1)$ for $d=6$, 7 and 8, respectively.
As expected, $A_D(d)$ monotonically decreases with increasing dimension, and are remarkably consistent with the expected relation $A_D = A_\Delta/3$ for $d=7$ and 8, although not for $d=6$.
The discrepancy in $d=6$ may be caused either by the large pre-asymptotic corrections, as suggested by Fig.~\ref{fig:msdpp}(b, inset)---the local slope does not reach the expected value given by the caging side---or by the more subtle critical scaling at $d_\mathrm{u}$. 

\subsection{Logarithmic Correction for $d=6$}

\begin{figure}
  \label{fig:corrcheck}
  \includegraphics[width=0.48\textwidth]{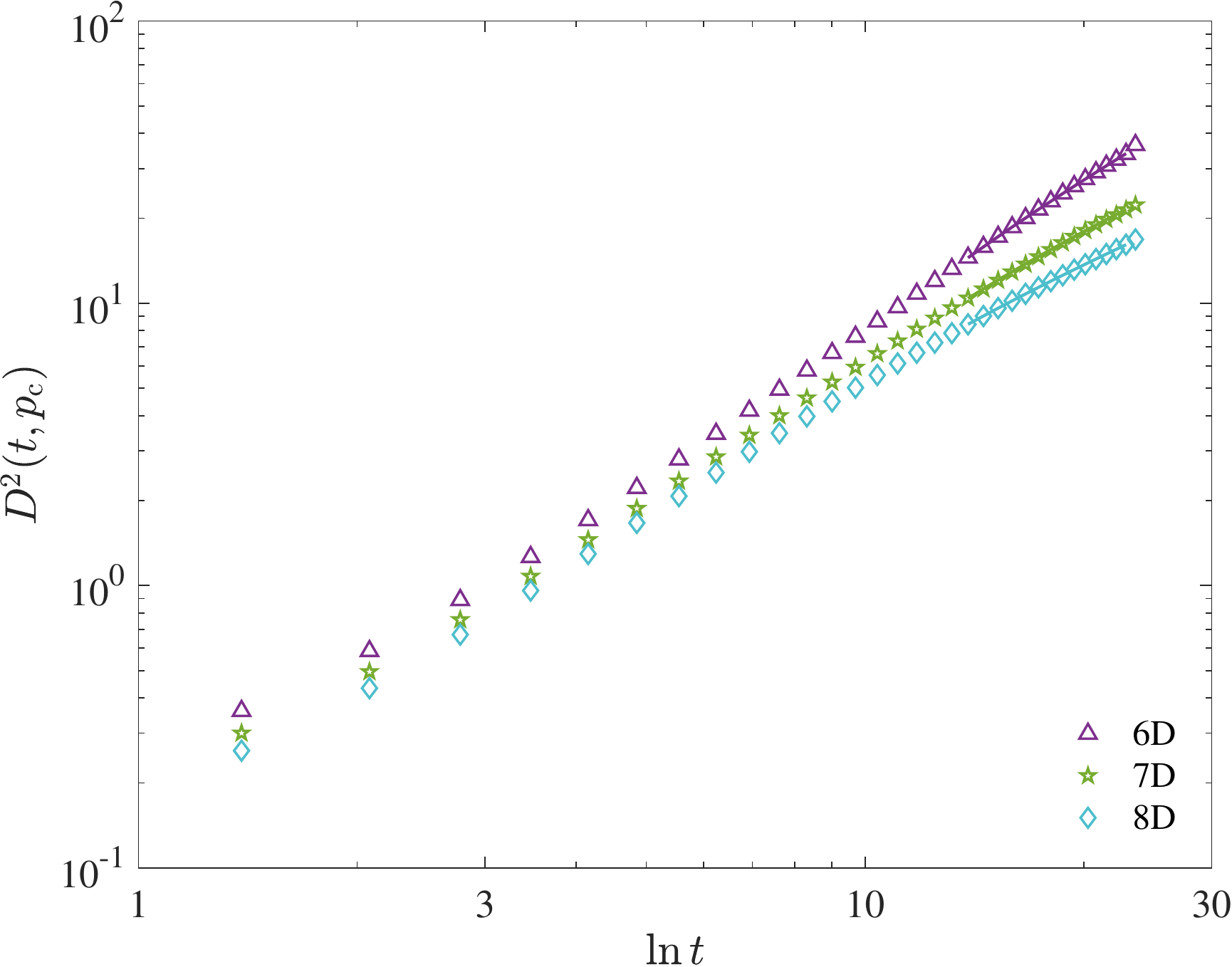}
  \caption{Growth of $\Delta^2$ with $\ln t$ at $p_\mathrm{c}$ for $d=6$. For comparison, results for $d=7$ and 8 as well as the fitting to Eq.~\eqref{eq:6dLogscaling} are plotted on the same scale.}
\end{figure}

We have thus far assumed that discrepancies from logarithmic scaling in $d=6$ were due to large pre-asymptotic corrections. We shall now examine the possibility that a logarithmic correction to the logarithmic scaling might be applied to exponent $\tau$. By explicitly presenting the constant term in equation~\eqref{eq:6dDscaling} as
\begin{equation*}
D^2(t,p_\mathrm{c}) = A_D (\ln t)^{1+\theta} + B, 
\end{equation*}
and taking the logarithm on both side gives
\begin{equation} \label{eq:6dLogscaling}
\log D^2 = (1+\theta) \ln \ln s + \ln A_D + \ln(1 + \frac{B}{A_D (\ln s)^{1+\theta}}). 
\end{equation}
Fitting this nonlinear equation, however, gives results that are sensitive to the number of short time data we discard (as observed in Ref.~\cite{mertens2018percolation}). For instance, we get $\theta=0.58(4)$ by fitting the results for $t \ge 2^{15}$ or $0.40(4)$ for $t \ge 2^{20}$, both of which are inconsistent with Ref.~\cite{mertens2018percolation} and the theoretical prediction $\theta=2/7$. This inconsistency might be the result of the large pre-asymptomatic corrections in $d=6$. 
It is also not possible determine whether or not $\theta$ is actually $0$, which would correspond to a simple logarithmic scaling. By comparison, similar fitting gives $\theta(d=7)=0.06(10)$ and $\theta(d=8)=0.003(3)$, both of which are consistent with the expected simple logarithmic scaling. Note that attempting to extract $\theta$ from the caging side dubiously gives $\theta \approx 1$, which likely reflects even more significant pre-asymptotic behaviors on this quantity. In short, the current results show only limited evidence in support of $\theta \neq 0$.

\subsection{Rescaled Mean-square displacement}

As a final test of the above critical exponents, we explicitly consider the scaling collapse discussed in Sect.~\ref{sec:theory} for $d\leq d_\mathrm{u}$. Figure~\ref{fig:msddyn}(a) shows that the mean-square displacement qualitatively exhibits the trend caging-subdiffusion-diffusion, as $p$ traverses the critical regime. The long time dynamics is either caged or diffusive, separated by the subdiffusion line at $p_\mathrm{c}$. The subdiffusion behavior is also observable at intermediate times for $p$ around $p_\mathrm{c}$. Note also that at short times the slope of all dynamical curves tends to $1$ with an intercept at $D^2(t=1) = p$, which is characteristic of the blind ant tracer dynamics. 

Rescaling time and $D^2$ as
\begin{align}
\tilde{t} = |p_\mathrm{c}-p|^{\mu+\mu_-} t \\
\tilde{D^2} = |p_\mathrm{c}-p|^{\mu_-} D^2
\end{align}
further reveals the quality of the expected collapse, shown in Figure~\ref{fig:msddyn}(b). While in $d=3$ and $4$, a reasonably good collapse can be obtained, for $d=5$ no neat collapse emerges over the accessible dynamical regime. This relatively poor scaling reflects the large pre-asymptotic corrections to scaling in both the caging and the subdiffusive regimes. For $d \ge 6$, the power-law scaling of $\Delta^2$ is replaced by a logarithmic growth and a collapse is, therefore, not expected for any $\epsilon < 0$, as observed. 
 
\section{Conclusion} \label{sec:conclusion}
We have investigated the mean-square displacement of an ant in a labyrinth on hypercubic lattices in $d=3$ to 8. For $d < d_\mathrm{u} = 6$, the expected power-law scalings of $\mu_-$ and $d'_\mathrm{w}$ are observed in simulations. 
Although pre-asymptotic corrections partially obfuscate the measurement of $d'_\mathrm{w}$ in $d=5$, in $d=3$ and $4$, 
we manage to obtain critical exponents that are almost an order of magnitude more accurate than previous estimates.
For $d \ge 6$, we derive the logarithmic scaling of both caging and subdiffusion, as well as the relation between their prefactors from the scaling analysis, which we validate in simulations. By our explicit consideration of Bethe lattices and random graphs, we confirm that this logarithmic growth persists in the limit $d\rightarrow\infty$, because its prefactor is nonvanishing. The ant in a labyrinth thus never undergoes a glass-like caging transition. Whether that is true for off-lattice percolation processes, which would be more akin to the Mari-Kurchan model, however, remains to be studied.

\begin{acknowledgments}
This research was supported by a grant from the Simons Foundation (\#454935, Giulio Biroli and \#454937, Patrick Charbonneau).
This research was also supported in part by the National Science Foundation under Grant No. NSF PHY17-48958.
The computations were carried out on the Duke Compute Cluster and Open Science Grid.
\end{acknowledgments}

\bibliography{abbrev}




\end{document}